# The Statistical Analysis of stars with Hα emission in IC 348 Cluster.

E. H. Nikoghosyan, A. V. Vardanyan, K. G. Khachatryan

*Byurakan Astrophysical observatory, Armenia, elena@bao.sci.am*

***Abstract.*** In this work the results of the statistical analysis of the ~ 200 stars with Hα emission in the IC 348 cluster are presented. The sample is completed up to R < 20.0. The optical radius is ~11′. The percentage of emission stars increases from bright to fainter objects and to the range of $13.0 \leq R-A_R \leq 19.0$ reaches 80%. The ratio between WTTau and CTTau objects is 64% and 36%. The 70% of X-ray sources are WTTau stars. The age of WTTau and CTTau objects are ~$2 \cdot 10^6$ years. The age of the non emission stars with a mass less solar is ~$2 \cdot 10^6$ years also, but non emission more massive objects are "older", the age of them is ~$7 \cdot 10^6$ years. The most massive stars with a low level of activity is concentrated in a small dense central core of the cluster with a radius ~1′, and apparently, they are generated during an earlier wave of star formation in the cluster.

***1. Introduction.*** The PMS stars are characterized by a number of observational features that appear on a very wide range of electromagnetic radiation from X-ray to radio emission. The quantitative ratio of these features changes during the evolution from a very young protostars to the evolution class III. Therefore, young star clusters, in which are simultaneously present stars formed during the successive waves of star formation, and therefore, at different stages of evolution, are natural laboratories for the study of the evolution of stars' observational characteristics.

The subject of our study is Hα stellar activity in the small young cluster IC 348, which is located at a distance ~300 pc on the eastern edge of the molecular cloud in Perseus. This cluster has always been a subject of active research (Herbig, 1998, Flaherty et al, 2013, Luhman et al, 2003, Herbst, 2008 and paper herein), hence, comprehensive observational data in X-ray, optical and infrared ranges have been obtained. The fundamental work on the study of stars with Hα emission belongs G. Herbig (Herbig, 1998, further H1998), in which are were revealed ~ 100 emission stars. In the future, their number has doubled (Luhman et al, 2003).

The aim of this work is the statistical analysis of the new, more completed sample of stars with Hα emission.

***2. Observation and data processing.*** The images of the study area were obtained in the primary focus of the 2.6 m telescope of the Byurakan Observatory, with the camera

SCORPIO and CCD with 2063 x 2059 pixels. The field and resolution of the images are 14′x14′ and 0.42″/pix respectively. The FWHM is not more than 2.5″.

The photometric observations with the filter R (Cousin) were held 22.01.2009 and 600 sec exposure. As photometric standards were used magnitudes of stars in NGC 7790 cluster.

The initial processing was performed according to standard procedure. The magnitudes were determined by using IRAF program. The errors of magnitudes are less than 0.04.

The search of star with Hα emission was carried out by slit-less method with grism and narrow-band interferences filter Hα ($\lambda c$ = 6560Å и $\Delta\lambda$ = 85Å). The observations were carried out in two epochs 20.01.2009 and 06.11.2009 with 2400 sec exposure. In first epoch spectral dispersion was 1.2 Å/pix, in second - 2.1 Å/pix. The equivalent widths EW(Hα) of Hα were determined by MIDAS program. The errors were determined by a formula adopted from (Vollmann & Eversberg, 2006): $\sigma(W_\lambda) = \sqrt{1 + \overline{Fc}/\overline{F}}\, (\Delta\lambda - W_\lambda)/(S/N)$, were $\overline{Fc}$ - is the average level of continuum, $\overline{F}$ - is the flux in the spectral line and S/N – is the signal/noise ration. For objects with R < 17.0 the errors of EW(Hα) not more than 30%, for weaker objects error increases up to 40% of the EW(Hα).

## 3. Results.
*3.1. The stars with Hα emission.* The list of the members of IC 348 cluster was taken from Flaherty et al, 2013 (F20013), which, in the area shown in Fig. 1, is completed up to I ≤ 22 (L2003). In this area was found about 200 stars with Hα emission, which equivalent widths, determined by the low-dispersion spectroscopy in Luhman et al, 2003 (L2003), are presented in the catalog of the cluster members (F2013). Previously, the EW(Hα) of 110 stars from this list have been already determined (H1998). In this case the spectral observations were carried out with higher dispersion (1.55 and 3.6 Å / pix), in our work, in cases where it was considered, have been used EW(Hα) taken from H1998.

In addition, we determined EW(Hα) for 16 stars. Their data, namely: coordinates, EW(Hα) and serial numbers in catalogs from F2013 and L2003 are presented in Table 1 and position are marked on Fig.1. For seven, marked with an asterisk objects (see Table 1), the emission was detected for the first time. The objects 5* was not included in the list of members in (F2013) and (L2003), but it was determined as probable member in Muench et al, 2007 (CXOANC J034421.6+321511, object 185 в Table 7). Note also, that for two objects, namely 6 and 12 (LkHa94), previously Hα absorption was not detected (H1998). In our case - they found a faint emission. This discrepancy is most likely the result of variability, which has been mentioned in H1998. Apparently, the variability of Hα emission is due to a significant discrepancy between the equivalent width in L2003 and H1998. In some cases the value of EW(Hα) differs by more than three times, and this fact can hardly be explained by errors of measurements.

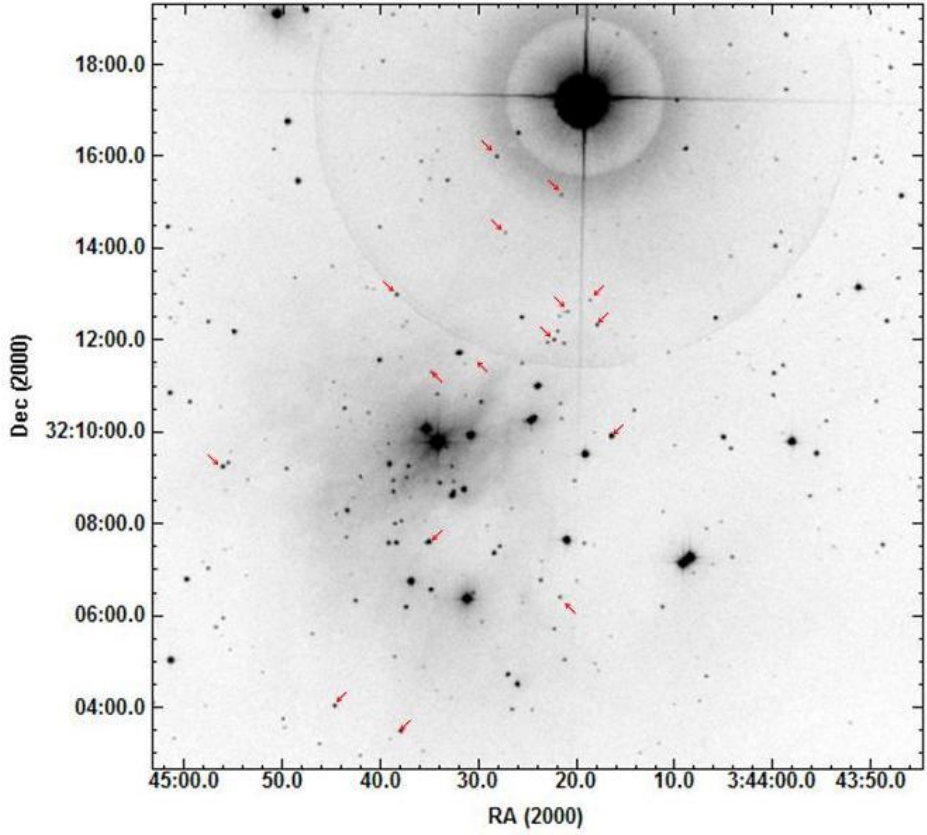

**Fig 1.** The field of study (DSS R2). The arrows show the position of emission stars from Table 1.

*Table 1*

## The Stars with Hα Emission

| N   | RA (2000)                              | Dec (2000)        | EW(Hα)   | F2013, L2003 |
|-----|----------------------------------------|-------------------|----------|--------------|
| 1*  | $3^h\ 43^m\ 58.57^s$                   | 32° 17′ 27.7″     | no cont  | 66, -        |
| 2*  | 3  44  16.44                           | 32  09  55.4      | 2        | 50, 50       |
| 3   | 3  44  17.88                           | 32  12  20.7      | 20       | 84, 83       |
| 4   | 3  44  21.32                           | 32  12  37.5      | 31       | 176, 167     |
| 5*  | 3  44  21.57                           | 32  15  10.0      | 39       | -, -         |
| 6   | 3  44  21.62                           | 32  06  24.6      | 2        | 109, 107     |
| 7   | 3  44  22.35                           | 32  12  01.2      | 75       | 92, 91       |
| 8*  | 3  44  27.22                           | 32  14  21.1      | 35       | 113, -       |
| 9   | 3  44  28.09                           | 32  16  00.5      | 3        | 77, 76       |
| 10* | 3  44  34.13                           | 32  16  36.2      | no cont  | 213, -       |
| 11* | 3  44  34.78                           | 32  11  18.4      | 4        | 129, 126     |
| 12  | 3  44  35.02                           | 32  07  35.9      | 2        | 23, 23       |
| 13* | 3  44  37.97                           | 32  03  29.7      | 3        | 34, 34       |
| 14  | 3  44  38.34                           | 32  12  59.6      | 2        | 123, 120     |
| 15  | 3  44  44.70                           | 32  04  02.5      | 46       | 15, 15       |
| 16  | 3  44  56.10                           | 32  09  15.1      | 11       | 20, 20       |

Totally we have identified 359 members of the cluster in this area and for 215 members an emission was detected. For two stars because of the weakness of the continuum we could not measure the equivalent widths. In the future, they will be considered as CTTau stars. In four cases, the emission was determined for the double object as a whole.

For a manning the list of the members in respect to R magnitude except the photometric data given in the F2013 and L2003 additionally they have been borrowed from a number of works: H1998; Cohen et al, 2004; Cieza & Baliber, 2006; Cieza et al, 2007. Besides, we identified R magnitudes for 23 stars (see Table 2).

*Table 2*

## R Magnitudes

| N  | RA (2000)                           | Dec (2000) | Rc    | F2013, L2003 |
|----|-------------------------------------|------------|-------|--------------|
| 1  | $3^h\ 43^m\ 46.4^s$                 | 32 11 06   | 19.61 | 271, -       |
| 2  | 3 43 47.6                           | 32 09 03   | 18.98 | 321, -       |
| 3  | 3 43 48.8                           | 32 15 51   | 17.38 | 137, -       |
| 4  | 3 43 49.5                           | 32 10 40   | 15.97 | 126, 123     |
| 5  | 3 43 51.2                           | 32 13 09   | 12.48 | 21, 21       |
| 6  | 3 43 59.7                           | 32 14 03   | 15.44 | 79, 78       |
| 7  | 3 44 00.5                           | 32 04 33   | 18.86 | 334, -       |
| 8  | 3 44 04.2                           | 32 13 50   | 16.81 | 112, 110     |
| 9  | 3 44 04.4                           | 32 04 54   | 18.23 | 231, 210     |
| 10 | 3 44 06.0                           | 32 15 32   | 18.46 | 242, -       |
| 11 | 3 44 06.2                           | 32 07 07   | 17.30 | 159, 153     |
| 12 | 3 44 08.9                           | 32 16 11   | 14.83 | 41, 41       |
| 13 | 3 44 11.6                           | 32 03 13   | 17.94 | 43, 43       |
| 14 | 3 44 19.0                           | 32 07 36   | 16.94 | 349, -       |
| 15 | 3 44 26.6                           | 32 08 21   | 17.34 | 73, 72       |
| 16 | 3 44 27.2                           | 32 14 21   | 16.53 | 133.-        |
| 17 | 3 44 21.6                           | 32 15 10   | 16.87 | M2007 185*   |
| 18 | 3 44 30.4                           | 32 09 45   | 18.02 | 289, 260     |
| 19 | 3 44 31.4                           | 32 16 36   | 18.43 | 213, -       |
| 20 | 3 44 34.7                           | 32 16 00   | 18.70 | 210, -       |
| 21 | 3 44 35.0                           | 32 15 31   | 17.24 | 152, -       |
| 22 | 3 44 35.9                           | 32 15 53   | 16.59 | 154, -       |
| 23 | 3 44 56.1                           | 32 09 15   | 14.95 | 20, 20       |

Totally, R magnitudes determined for 263 objects in general and for the 194 stars with Hα emission. The total list of the objects with serial numbers from F2013, L2003, coordinates, R magnitudes, $A_v$, EW(Hα), Sp, $L_{bol}$, $T_{eff}$, Lx и α (SED slope) are presented in the Appendix.

***3.2. The statistical analysis of the stars with Hα emission.*** One of the main tasks of this paper is to estimate the percentage of objects with Hα emission in respect to PMS stars in the cluster. Fig. 2 shows two luminosity functions for all objects and stars with emission in the visible R mag (left) and corrected by the absorption (right). The $A_V$ values adopted from L2003 and $A_R/A_V$ = 0.748 ratio from (Rieke, M. J. Lebofsky, 1985). The histogram on the left side clearly shows that the percentage of stars with Hα emission varies, depending on the range of magnitudes. It is also well reflected in the graph presented on Fig. 3, which shows the distribution of the fraction of emission stars from all objects with R mag in this range. As shown in Fig. 3, the proportion of emission stars in the range $14.0 \leq R \leq 20.0$ reaches ~ 80% and remains practically unchanged. Among the bright stars with R ≤ 13.0 no objects with Hα emission, and among faint stars (R ≥ 20.0) the percentage of emission stars is lower than in $14.0 \leq R \leq 20.0$ range. The decrease in the percentage of the emission objects among the fainter stars in the first, of course, can be explained by incompleteness of the observational data. The cluster members up to R = 20.0 are practically fully complicated for stars with Hα emission.

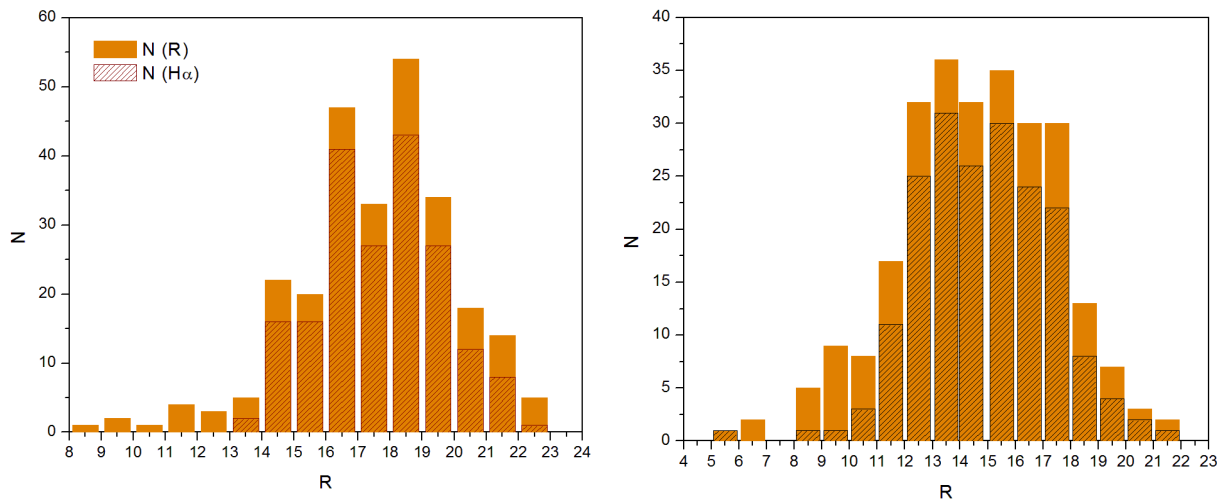

**Fig. 2.** R luminosity function for visible magnitudes (left) and corrected for the absorption (right).

However, to represent the real picture of Hα activity of the cluster's stellar population is necessary to estimate the percentage of stars with Hα emission depending not on the visible but the real magnitudes. Indeed, taking into account the absorption introduced certain adjustments. On one magnitude shifted to the left the range with a relatively constant fraction of emission stars. In this case, it corresponds to the range of $13.0 \leq R \leq 19.0$ (see. Fig. 3). The percentage of emission stars among the relatively bright objects (R ≤ 13.0) is less: from 10% to 60%, increasing with increase in magnitude. Thus, we can conclude that Hα activity of the young stellar population in IC 348 cluster depends on the absolute brightness of the objects, from the brightest to the fainter, the fraction of the emission stars increases, reaching ~80% in the range of 13.0 ≤ R ≤ 19.0.

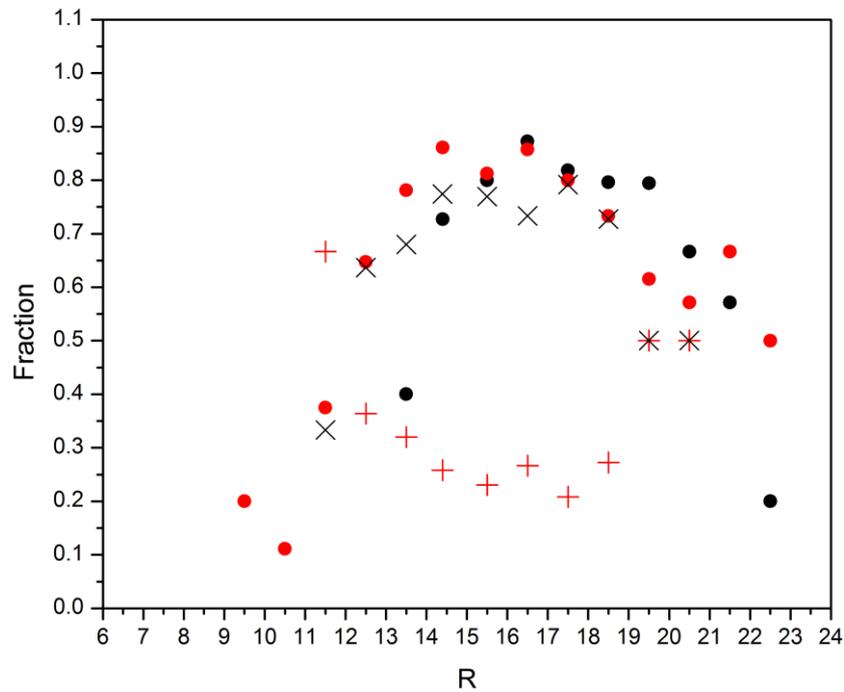

**Fig. 3.** The fraction of stars with emission Hα: black circles - respect to visible R magnitude, red circles - respect to R - $A_R$, black crosses - the fraction of WTTau on all Hα emission stars, red crosses - the fraction of CTTau.

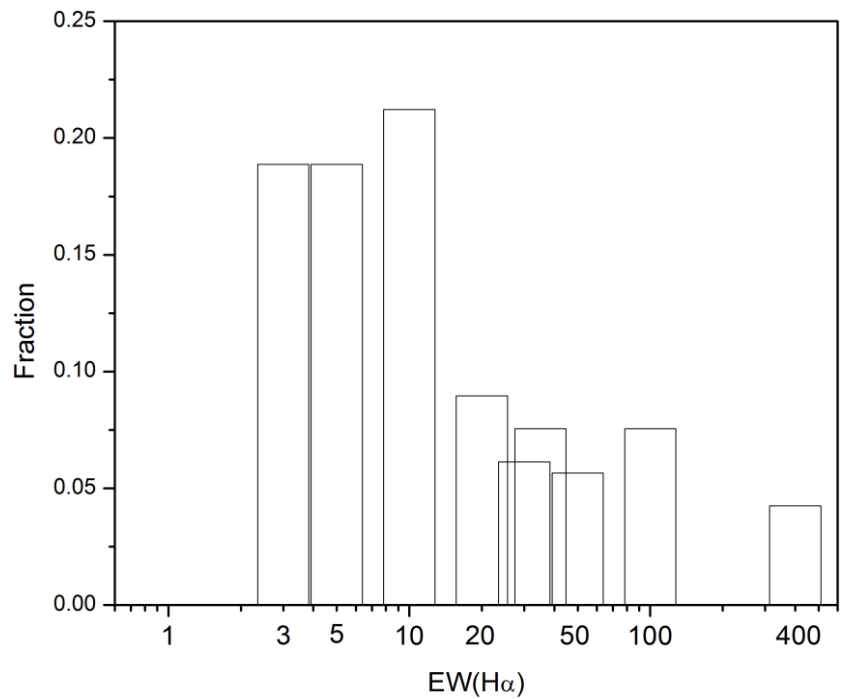

**Fig. 4.** The relative content of emission stars, depending on the EW (Hα).

Even in H1998 drew attention to the fact that in the cluster IC 348, compared with other young star clusters, abnormally high levels of WTTau objects. Increase in the number of identified stars with Hα emission (from 110 to 215) doesn't make significant changes, which is reflected in the diagram shown in Fig. 4. About 60% of all emission stars have EW(Hα) ≤ 10 Å, which considers with the result obtained in H1998.

For the classification of emission stars relative to the equivalent width, we have used criteria set out in White & Basri, 2003, according to which, the limit value of EW(Hα) for WTTau and CTTau objects depends on their spectral class and is 3 Å for Sp earlier than K5, 10 Å - for objects with Sp from K6 to M3 and 20 Å - for Sp latter than M3. According to this classification the percentage of WTTau и CTTau objects are 64% and 36% respectively (see Table 3). Moreover, the ratio between the CTTau and WTTau stars changes depending on the brightness. In the range 13.0 ≤ R ≤ 19.0 WTTau/CTTau ≈ 3/1, and this ratio reduces for both brighter and fainter stars (see Fig. 3). The increase in the fraction of CTTau stars with R > 19.0 is likely arises, because of the selectivity of the observational data - for faint objects is much easier to identify a strong emission than weak. Among bright stars the common percentage of emission objects is very small. Therefore, to do any conclusions in this regard is not advisable.

The WTTau/CTTau relation also varies depending on the spectral classes (see Table 3): for stars with Sp from K6 to M3 the percentage of WTTau is 59% and for objects with Sp latter than M3 is already 68%. Furthermore, it should be noted that from early to later spectral classes the percentage of emission stars generally increases: the stars with early Sp contain only 20% emission star, while the objects with Sp later than M3 - is 85%.

*Table 3*

**Relation Between Hα Activity and Other Parameters**

| Selection (N) | WTTau | CTTau |
|---|---|---|
| Hα emission (214) | 138 | 77 |
| Earlier than K5 (34) | 4 | 3 |
| K6 - M3 (84) | 36 | 25 |
| Later than M3 (164) | 95 | 44 |
| -2.56 < α < -1.8 (Class III) (182) | 102 | 8 |
| -1.8 < α < 0 (Class II) (142) | 29 | 59 |
| α > 0 (Class I) (21) | - | 6 |
| Lx (153) | 89 | 25 |

Table 3 shows quantitative estimate of WTTau CTTau in relation to the other signs of activity of PMS star. One of them is infrared excess characterized by the slope of the spectral energy distribution ($\alpha$) in the mid-infrared range, and which is depend of the evolutionary class of PMS stars (Lada et al, 2006). According to the data given in Table 3, the slope of SED ($\alpha$) and intensity of H$\alpha$ emission are in a very good agreement, because among the stars with $-2.56 < \alpha < -1.8$ vast majority are WTTau objects. All 6 objects with highest infrared excess ($\alpha > 0$) in which revealed H$\alpha$ emission are CTTau stars. Among the Class II objects are dominated the stars with strong emission (67%).

According to the data in Table 3 it also clearly shows that among the stars with weak emission, i.e., Class III objects, the percentage of X-ray sources are significantly higher which very well agrees with the previous results obtained for the IC 348 (Stelzer et al, 2012) as well as in other clusters (Feigelson et al, 2006).

***3.3. The age.*** Fig. 5 shows the position of WTTau and CTTau objects, as well as the stars in which H$\alpha$ emission have not been revealed, concerning the isochrones constructed on the model proposed in D'Antona & Mazzitelli, 1994. Totally $T_{eff}$ and $L_{bol}$ are identified for 300 objects. The position of the stars on the diagram shows that the two classes of objects relative H$\alpha$ activity no significant difference either in weight or age. The position of non emission stars is markedly different. These objects can be divided in two groups. One of them includes the stars with masses less than solar and which positions on the diagram is not much different from the emission ones. Indeed, the median age for all three categories of the stellar population of the cluster is almost the same and equal to $\sim 2 \cdot 10^6$ years. Note, that for stellar objects with masses more than 0.4 $M_\odot$ we used the evolutionary model proposed in Baraffe et al, 1998. This estimation is slightly higher than the value obtained for the emission stars in the H1998. The second group of non emission stars includes stars with masses higher than solar. The median value of age, computing according to the model proposed in D'Antona & Mazzitelli, 1994, for these objects is $7 \cdot 10^6$ years, hence, they are older.

***3.4. The distribution of the stars in the cluster.*** We have also considered the question of the distribution of stars with different degrees of activity relative H$\alpha$ activity in the field of the cluster. Fig. 6 shows the radial distribution of the stellar density relative to the cluster center ($03^h44^m34^s$, $+ 32°09'48''$, Wu et al, 2009) for different samples of stellar objects, which have been determined for each ring with a width 0.5′ by simply dividing the number of stars in the surface area. The center of the cluster is associated with the brightest star HD 281159B. Measure of uncertainty determined according to Poisson statistics on the number of stars in each ring. On the Fig. 6a are presented a radial density distributions for stars with different degrees of H$\alpha$ activity: CTTau, WTTau and objects which are not detected an emission. For comparison, on Fig. 6b are presented the radial density distributions of stars to different Sp , namely latter than K6; later than K6 and earlier than M3, and finally star with Sp latter than M3. Fig. 6c shows a radial density distribution of stellar objects of different evolutionary classes defined on the basis of mid-IR infrared excess and as well as the X-ray sources.

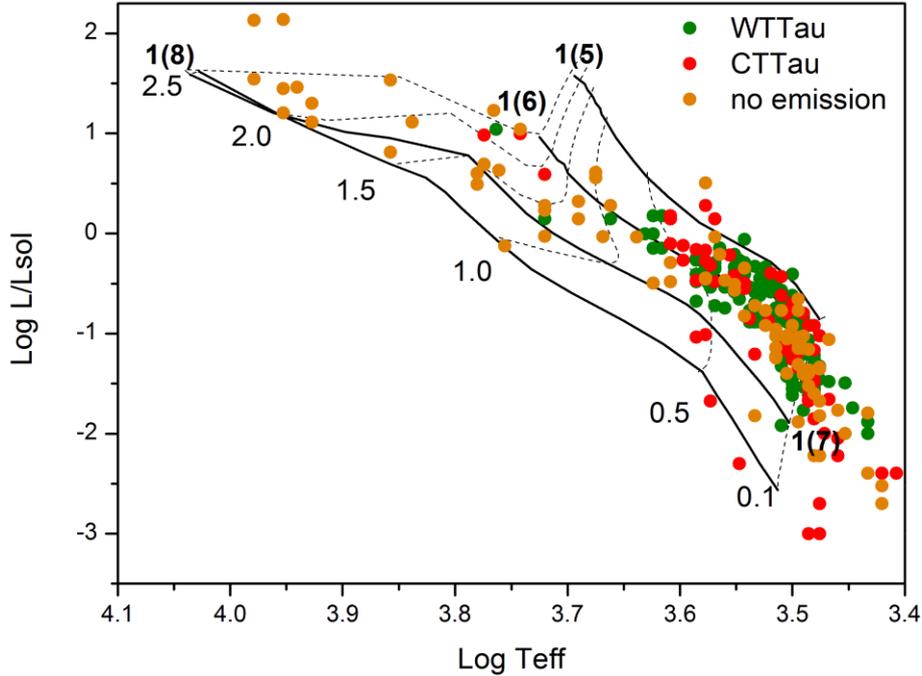

**Fig. 5.** Position of the stars with varying degrees of Hα activity relative isochrones (D'Antona & Mazzitelli, 1994)

The diagrams clearly show that, as was noted in the previous studies (Herbst, 2008 and works mentioned herein), in the center of the cluster is located the core of high density. Among almost all samples, except for stars with late spectral classes and Class I YSOs, was observed a pronounced tendency to concentrate in the core with a radius ~1′. However, objects of various samples distributed differently across the cluster's field. Non emission stars, mostly concentrated in a small central part with radius ~ 30″. Emission stars relatively uniformly distributed in the cluster that was more marked in H1998. However, they are also, (although not as well-defined) tend to concentrate towards the center and WTTau with more pronounced tendency. This agrees well with the distribution of stars with different evolutionary classes. Notes that in the central core there are not the youngest Class I YSOs. If we consider the distribution of objects with respect to their spectral class or masses, there is clearly seen the following picture - the more massive stellar objects are stronger concentrated in the central region. The most of X-ray sources are concentrated in the center of the cluster.

Thus we can conclude that in the central region of the cluster in which the stellar density is 20 times more than in the cluster field, concentrated more massive stellar objects with low Hα activity and a small infrared excess, i.e. the members of the cluster at a later stage of evolution. However, they also represent the largest amount of X-ray sources, which confirms that most of X-rays activity observed in young stars is a result of chromospheres' activity (Feigelson et al, 2006).

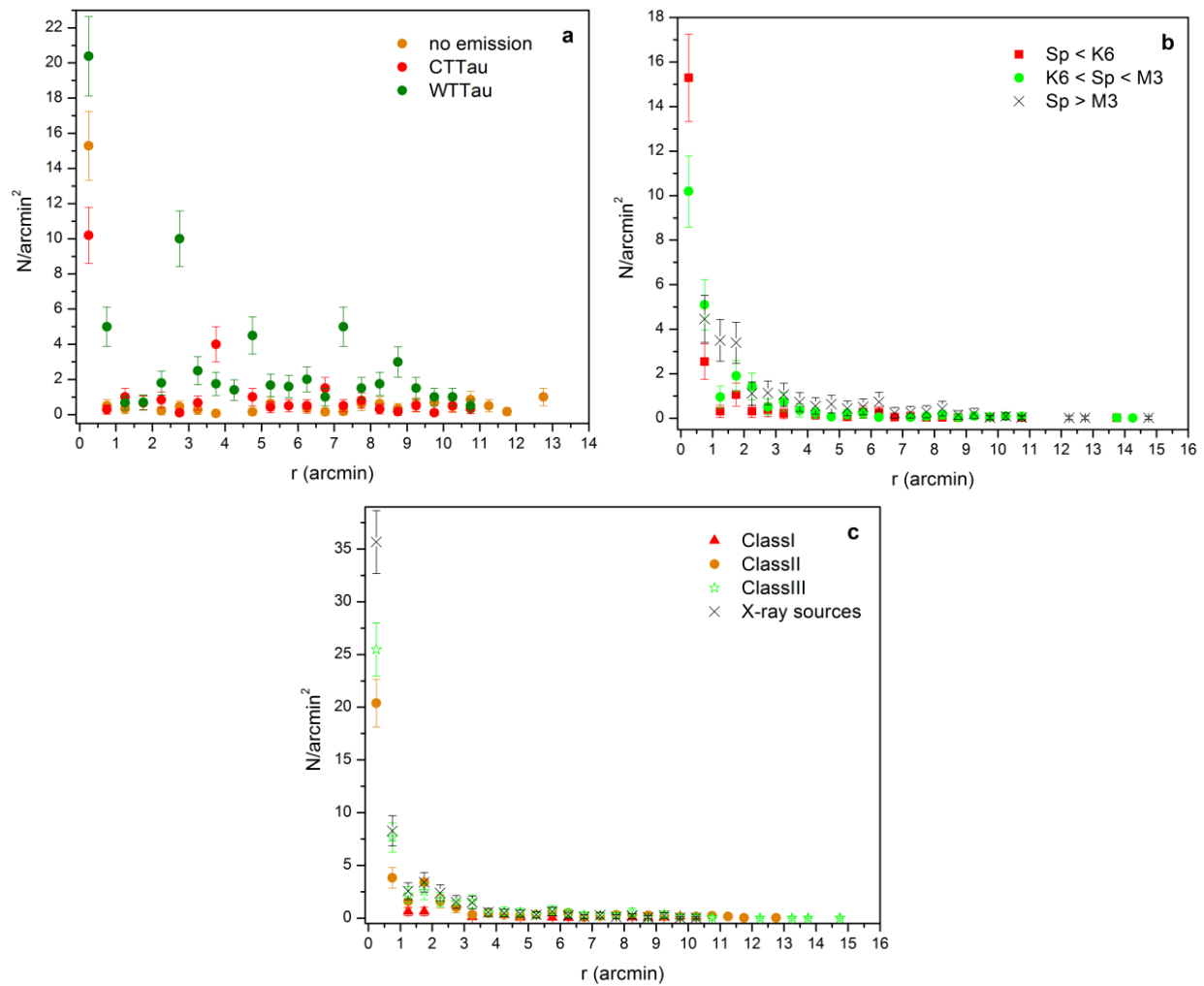

**Fig. 6.** Radial density distribution of stellar objects: relative Hα activity (a), spectral classes (b), infrared excess and X-ray (c).

*4. Discussion.* Active research of the young star cluster IC 348, allowed along with other properties also identified in it 215 stars with Hα emission. After perhaps the most fundamental work in this area H1998, in which was presented a list and a detailed study of 110 emission stars, their number almost doubled and reaches to 215. The number of objects so greatly increased primarily due to expansion of the field of observations in subsequent papers. If in H1998 for the value of the cluster's optical radius provides value 4.0′, but now, the farthest emission object located on a distance ~ 11′ from HD 281159B. The increase in the number of detected emission stars is also associated with a decrease in the threshold of the equivalent width - if in H1998 are only three stars with EW (Hα) <

3, but now such objects are ~ 20. On about one magnitude (from 19.0 to 20.0) increases the limit of the completeness of the stars with H$\alpha$ emission, that reflects the R luminosity function (see par. 3.2).

Accounting of an absorption showed that the percentage of emission stars on their brightness R band has a significant gradient - it increases from bright to the fainter and on the range $13.0 \leq R \leq 19.0$ reaches ~80%. Then the percentage of emission stars falls, which, of course, is primarily due to an incomplete of the observational data, and does not allow definite conclusions. Apparently, 80% is the lower limit estimate of the relative number of stars with H$\alpha$ activity. One possible reason for the lack of H$\alpha$ emission in the remaining 20% - a variability between low emission and absorption, as already noted in the par. 3.1., and that, taking into account the high percentage of stars with a small value of EW (H$\alpha$), can make a significant adjustment in the assessment of the percentage of emission objects. The second - the close binaries, which can also be the cause of variability, and finally - the measurement errors, which also cannot be excluded. Taking into account these factors certainly should raise the percentage of emission stars and approach it to 100%. The percentage of the emission stars among brighter and fainter objects is lower. In the first case - a reflection of reality, in the second - most likely the result of selection.

The distribution of the emission stars according to their spectral classes it is in good agreement with the above analysis. From early to later the percentage of emission objects is increasing: the only 20% of stars with early Sp has H$\alpha$ emissions, while objects with Sp later than M3 - already 85%.

As been already mentioned, there is anomalously high content of WTTau stars in the cluster IC 348, compared to the other star forming regions, such as the Tau-Aur (Kenyon et al, 2008), L1641 (Wouterloot & Brand, 1992), NGC 2264 (Dahm & Simon, 2005). Increasing number of objects does not change this relation. About 60% of emission stars has EW(H$\alpha$) < 10 (see Fig. 4). Moreover, the classification of WTTau and CTTau according to the criteria proposed by the authors of (White & Basri, 2003) shows, that the percentage of WTTau even rose slightly up to 64%. Note that in H1998 it was 58%. In this regard, it should be noted that practically the same quantitative relation is between these two classes of objects observed in the cluster Cep OB3, whose age is significantly less (~$0.7 \cdot 10^6$ year, Nikoghosyan, 2013). This ratio between these two types of emission stars, which essentially reflects their evolutionary status, is in good agreement with the classification according to SED slope ($\alpha$) in the infrared range. By this classification the "oldest" Class III objects with $-2.56 < \alpha < -1.8$ constitute ~50% and the Class II objects
($-1.8 < \alpha < 0$) is about 30% of the stellar population in the clusters (Lada et al, 2005). Moreover, there is a well-defined relationship between spectral class and value of "$\alpha$". For stars with Sp earlier than K6 only 11% of objects have optically thick disk, and among the objects with Sp from K6 to M2 - already 47% (Lada et al, 2005).

This fact agrees well with the data presented in Table. 3, according to which, among the disk-bearing stars (-2.56 < α < -1.8) the vast majority is also has weak emission, and among the objects with optically thick disk (-1.8 < α < 0) there is a larger percentage of CTTau objects. Among the youngest stellar population (Class I YSOs with α > 0) the percentage of emission stars is very small and they all have a strong emission. Presumably, Hα emission is strongly absorbed in a optically thick disks' matter and its intensity is below the threshold of sensitivity of the detectors.

Apparently, because of this absorption among the X-ray sources are mostly (70%) the WTTau stars. However, the X-ray emission is observed as well as in the stars with strong emission. If the X-ray emission in the first case is mainly a product of chromospheric bursts, then among the CTTau stars, it is caused by accretion activity which also assumes the presence of a massive disk. On the other hand, a massive disk can absorb a significant fraction of the X-rays. This, rather than an absence of X-ray activity, is presumably while the X-ray emission in a large number of the CTTau stars lies below the detection threshold in this range.

The members of the cluster are unevenly distributed in the field of the cluster. In the central region, the density which is more than 20 times greater than the average density of the cluster, concentrated early-type stars with a lower degree of emission activity and smaller infrared excess (see par. 3.4). Here also concentrated the majority of X-ray sources. On one hand, this is due to the fact that there are concentrated most massive stars with more strong X-rays emission, on the other - WTTau objects, in which X-ray radiation experienced significantly lower absorption. Furthermore, as shown in par. 3.3, concentrated in the center of the cluster massive stars are "older." The median value of their age is $\sim 7 \cdot 10^6$ years. The age of other stellar objects including star with weak and strong emission, as well as non emission stars with masses lass than solar is $\sim 2 \cdot 10^6$ years.

Thus, it can be concluded that the star formation process in the cluster has at least two stages and located in the center of the cluster relatively more massive stars, which currently have a low level of activity were likely formed during the earlier stage of star formation.

*5. Conclusion.* The statistical analysis of ~200 stars with Hα emission revealed in the cluster IC 348 lead to the following conclusions:
- The sample of emission stars is completed to R < 20.0 mag.;
- The optical radius of the cluster is ~11′;
- The percentage of emission stars increases from bright to fainter objects and to the range $13.0 \leq R-A_R \leq 19.0$ reaches 80%, which is probably the lower limit;

- The percentage of emission stars increases with decreasing their spectral classes;
- The ratio between WTTau and CTTau objects is 64% and 36% respectively, which correlates well with the ratio of objects with different evolutionary stage determined according to their infrared excess;
- The main part of X-ray sources (~70%) are WTTau objects;
- The age of WTTau and CTTau stars are ~2·10$^6$ years;
- The age of non emission stars with masses less than solar is also ~2·10$^6$ years, the age of the brightest objects is ~7·10$^6$ years;
- The most massive stars with a low level of activity are concentrated in a small dense central core of the cluster with a radius ~1′.

This work was supported by grant No. 13-1S087 of the State Committee on Science of the Republic of Armenia

# Appendix

## List of Objects

| N (F2013) | N (L2003) | RA (2000) | | | Dec (2000) | | | R | W(Ha) | Teff | Lum | Av | Lx | alfa | Sp |
|---|---|---|---|---|---|---|---|---|---|---|---|---|---|---|---|
| 85 | 84 | 3 | 43 | 32.1 | 32 | 6 | 17 | 14.81 | 1 | 3741 | 0.29 | 0.96 | | -2.81 | M0.75 |
| 168 | 161 | 3 | 43 | 33.6 | 32 | 1 | 45 | 17.93 | 5 | 3125 | 0.1 | 2.34 | | -2.9 | M5 |
| 369 | | 3 | 43 | 44.3 | 32 | 3 | 42 | | | | | 0 | | -1.07 | |
| 63 | 63 | 3 | 43 | 44.6 | 32 | 8 | 18 | | 35 | 3741 | 0.48 | 2.41 | | -2 | M0.75 |
| 197 | | 3 | 43 | 45.2 | 32 | 3 | 59 | | | | | 0 | | -0.03 | |
| 271 | | 3 | 43 | 46.4 | 32 | 11 | 6 | 19.61 | | 2935 | | 0 | | -1.4 | |
| 321 | | 3 | 43 | 47.6 | 32 | 9 | 3 | 18.98 | | 2880 | 0.017 | 1 | | -1.22 | |
| 208 | 193 | 3 | 43 | 48.6 | 32 | 13 | 51 | 17.92 | 6 | 3125 | 0.06 | 0.99 | | -2.62 | M5 |
| 98 | 97 | 3 | 43 | 48.8 | 32 | 7 | 33 | 15.44 | | 3632 | 0.34 | 2.62 | | -2.74 | M1,5 |
| 137 | | 3 | 43 | 48.8 | 32 | 15 | 51 | 17.38 | | 3197 | | 0 | | -0.77 | |
| 126 | 123 | 3 | 43 | 49.4 | 32 | 10 | 40 | 15.97 | 11 | 3342 | 0.17 | 1.21 | 34.2 | -2.67 | M3,5 |
| 325 | | 3 | 43 | 50.6 | 32 | 3 | 18 | | | 2477 | | 0 | | -0.75 | G5 |
| 370 | | 3 | 43 | 51 | 32 | 3 | 8 | | | | | 0 | | -0.73 | |
| 371 | | 3 | 43 | 51 | 32 | 3 | 26 | | | | | 0 | | -0.73 | |
| 21 | 21 | 3 | 43 | 51.2 | 32 | 13 | 9 | 12.48 | | 5770 | 4.3 | 2.62 | 472 | -2.73 | |
| 359 | | 3 | 43 | 51.6 | 32 | 12 | 40 | | | | | 0 | | 0.43 | |
| 202 | 188 | 3 | 43 | 53.8 | 32 | 7 | 30 | 18.04 | 4 | 3234 | 0.09 | 2.8 | | -2.66 | M4,25 |
| 108 | 106 | 3 | 43 | 54.6 | 32 | 0 | 30 | 16.91 | 7 | 3234 | 0.27 | 3.05 | | -2.61 | M4,25 |
| 234 | 213 | 3 | 43 | 55.1 | 32 | 7 | 15 | 19.12 | | 2990 | 0.046 | 1.45 | | -2.97 | M6 |
| 204 | 190 | 3 | 43 | 55.3 | 32 | 7 | 53 | 18.49 | 40 | 3024 | 0.068 | 1.28 | | -0.98 | M5,75 |
| 44 | 44 | 3 | 43 | 55.5 | 32 | 9 | 32 | 14.03 | | 5250 | 1.9 | 3.76 | 639.3 | -2.83 | K0 |
| 26 | 26 | 3 | 43 | 56 | 32 | 2 | 13 | | 44 | 4060 | 1.5 | 7.34 | | -0.88 | K7 |
| 122 | 119 | 3 | 43 | 56.2 | 32 | 8 | 36 | 15.95 | 4 | 3850 | 0.33 | 2.94 | 46.5 | -2.86 | M0 |

| | | | | | | | | | | | | | | |
|---|---|---|---|---|---|---|---|---|---|---|---|---|---|---|
| 365 | | 3 | 43 | 56.2 | 32 | 3 | 6 | | | | | 0 | | -0.13 | |
| 279 | 250 | 3 | 43 | 56.4 | 32 | 9 | 59 | 21.28 | | 2838 | 0.01 | 1.28 | | -2.92 | M7,25 |
| 377 | | 3 | 43 | 56.9 | 32 | 3 | 3 | | | | | 0 | | 0.42 | |
| 167 | | 3 | 43 | 57.2 | 32 | 1 | 34 | | | 2935 | 0.087 | 9.2 | | -1.28 | |
| 46 | 46 | 3 | 43 | 57.6 | 32 | 1 | 37 | 22.05 | | 3778 | 0.36 | 10.21 | | -2.33 | M0,5 |
| 66 | | 3 | 43 | 58.6 | 32 | 17 | 28 | | no cont | 3342 | | 0 | | -1.18 | |
| 59 | 59 | 3 | 43 | 58.9 | 32 | 11 | 27 | 16.09 | 28 | 3596 | 0.61 | 3.9 | | -1.42 | M1,75 |
| 103 | | 3 | 43 | 59.1 | 32 | 14 | 21 | 20 | 14 | 3342 | | 0 | | -1.07 | |
| 262 | 237 | 3 | 43 | 59.2 | 32 | 5 | 57 | 19.68 | 8 | 3024 | 0.029 | 1.84 | | -2.33 | M5,75 |
| 364 | | 3 | 43 | 59.2 | 32 | 2 | 51 | | | 2990 | 0.021 | 0 | | 0.11 | |
| 372 | | 3 | 43 | 59.4 | 32 | 0 | 35 | | | | | 0 | | -0.61 | |
| 14 | 14 | 3 | 43 | 59.6 | 32 | 1 | 54 | | | 3778 | 3.2 | 14.65 | | -0.33 | M0,5 |
| 79 | 78 | 3 | 43 | 59.7 | 32 | 14 | 3 | 15.44 | 1.5 | 3741 | 0.33 | 1.67 | 820.6 | -2.76 | M0,75 |
| 223 | 203 | 3 | 43 | 59.9 | 32 | 4 | 42 | 19.08 | 13 | 3024 | 0.048 | 1.74 | | -1.42 | M5,75 |
| 334 | | 3 | 44 | 0.5 | 32 | 4 | 33 | 18.86 | | 3125 | | 0 | | -0.86 | |
| 268 | 242 | 3 | 44 | 2.3 | 32 | 10 | 16 | 20.02 | 7 | 3091 | 0.029 | 3.09 | | -2.64 | M5,25 |
| 376 | | 3 | 44 | 2.4 | 32 | 1 | 40 | | | | | 0 | | 0.42 | |
| 373 | | 3 | 44 | 2.4 | 32 | 2 | 5 | | | | | 0 | | 1.33 | |
| 136 | 133 | 3 | 44 | 2.6 | 32 | 1 | 35 | 16.98 | 6 | 3161 | 0.14 | 1.21 | 7.3 | -2.79 | M4,75 |
| 374 | | 3 | 44 | 2.6 | 32 | 2 | 0 | | | | | 0 | | 1.01 | |
| 206 | 192 | 3 | 44 | 3.6 | 32 | 2 | 34 | 17.96 | 11 | 3125 | 0.04 | 1.6 | 1.3 | -2.8 | M5 |
| 207 | 191 | 3 | 44 | 3.6 | 32 | 2 | 34 | 17.96 | | | 0.04 | 1.84 | | -2.18 | |
| 147 | 143 | 3 | 44 | 4.1 | 32 | 7 | 17 | 16.56 | 3 | 3632 | 0.18 | 2.27 | 51.7 | -2.82 | M1,5 |
| 112 | 110 | 3 | 44 | 4.2 | 32 | 13 | 50 | 16.81 | 5 | 3161 | 0.22 | 2.34 | 9.2 | -2.65 | M4,75 |
| 231 | 210 | 3 | 44 | 4.4 | 32 | 4 | 54 | 18.23 | 9 | 3024 | 0.05 | 1.67 | 4.6 | -2.14 | M5,75 |
| 52 | 52 | 3 | 44 | 5 | 32 | 9 | 54 | 14.09 | | 4660 | 0.93 | 2.62 | 37.3 | -2.83 | K3,5 |
| 335 | | 3 | 44 | 5.8 | 32 | 0 | 1 | | | 3057 | | 0 | | -0.92 | |
| 333 | | 3 | 44 | 5.8 | 32 | 0 | 28 | | | | | 0 | | -0.49 | |
| 242 | | 3 | 44 | 6 | 32 | 15 | 32 | 18.46 | | 2935 | | 0 | | -1.18 | |
| 158 | 152 | 3 | 44 | 6.1 | 32 | 7 | 7 | 17.82 | 5.5 | 3234 | 0.15 | 4.65 | 10.9 | -2.3 | M4,25 |
| 159 | 153 | 3 | 44 | 6.2 | 32 | 7 | 7 | 17.3 | | | | 0 | | -2.3 | |
| 132 | 129 | 3 | 44 | 6.8 | 32 | 7 | 54 | 17.37 | 41 | 3234 | 0.17 | 2.7 | 5.6 | -1.09 | M4,25 |
| 344 | | 3 | 44 | 6.9 | 32 | 1 | 55 | | | | | 0 | | -0.34 | |
| 177 | 168 | 3 | 44 | 7.5 | 32 | 4 | 9 | 17.9 | 14 | 3161 | 0.099 | 2.09 | | -1.65 | M4,75 |
| 196 | 183 | 3 | 44 | 7.7 | 32 | 5 | 5 | 18.86 | 3 | 3198 | 0.089 | 3.9 | 7.6 | -2.46 | M4,5 |
| 7 | 7 | 3 | 44 | 8.5 | 32 | 7 | 17 | | | 9520 | 35 | 0.92 | | -2.7 | A0 |
| 41 | 41 | 3 | 44 | 8.9 | 32 | 16 | 11 | 14.83 | | 5250 | 1.7 | 3.55 | | -2.78 | K0 |
| 8 | 8 | 3 | 44 | 9.1 | 32 | 7 | 9 | 9.73 | | 8970 | 28 | 0.96 | | -2.38 | A2 |
| 215 | 196 | 3 | 44 | 9.2 | 32 | 2 | 38 | 20.83 | 35 | 3850 | 0.092 | 6.24 | 124.6 | 0.12 | M0 |
| 146 | 142 | 3 | 44 | 10.1 | 32 | 4 | 4 | 18.2 | 85 | 3024 | 0.12 | 2.02 | | -1.34 | M5,75 |
| 260 | 235 | 3 | 44 | 10.2 | 32 | 7 | 34 | 19.48 | 4 | 3024 | 0.028 | 1.42 | | -1.74 | M5,75 |
| 283 | 254 | 3 | 44 | 11.1 | 32 | 1 | 44 | | 400 | 2632 | 0.004 | 2.16 | | -0.56 | M8,25 |
| 138 | 134 | 3 | 44 | 11.2 | 32 | 8 | 16 | 17.37 | 5 | 3091 | 0.14 | 1.42 | 4.1 | -2.65 | M5,25 |
| 95 | 94 | 3 | 44 | 11.3 | 32 | 6 | 12 | 15.63 | 5 | 3850 | 0.39 | 2.7 | 55 | -2.83 | M0 |
| 117 | 114 | 3 | 44 | 11.4 | 32 | 19 | 40 | 16.13 | 10 | 3415 | 0.2 | 1.84 | | -2.73 | M3 |
| 43 | 43 | 3 | 44 | 11.6 | 32 | 3 | 13 | 17.94 | 4.5 | | | 0 | | -1.07 | ? |
| 255 | 232 | 3 | 44 | 12.8 | 32 | 10 | 55 | 18.86 | 6 | 3058 | 0.024 | 0.92 | 2.5 | -2.21 | M5,5 |
| 342 | | 3 | 44 | 12.9 | 32 | 13 | 24 | | | | | 0 | | -0.21 | |
| 48 | 48 | 3 | 44 | 13 | 32 | 1 | 35 | 21.26 | 45 | | | 0 | 156.8 | 0.25 | ? |
| 247 | 224 | 3 | 44 | 13 | 32 | 13 | 16 | 18.96 | | 3091 | 0.039 | 2.2 | | -1.71 | M5,25 |
| 116 | 113 | 3 | 44 | 13.6 | 32 | 15 | 54 | 17.7 | 5 | 3415 | 0.26 | 5.14 | 182.8 | -2.75 | M3 |
| 343 | | 3 | 44 | 14.9 | 32 | 13 | 43 | | | | | -1.77 | | -2.37 | |
| 218 | | 3 | 44 | 15.2 | 32 | 19 | 42 | | | 3160 | | 0 | | -1.46 | |
| 243 | 220 | 3 | 44 | 15.6 | 32 | 9 | 22 | 20.17 | 15 | 2795 | 0.018 | 0.57 | | -2.65 | M7,5 |
| 345 | 284 | 3 | 44 | 16.2 | 32 | 5 | 41 | | | 2400 | 0.002 | 1.6 | | -0.99 | M9 |
| 50 | 50 | 3 | 44 | 16.4 | 32 | 9 | 55 | 13.52 | 2 | 5250 | 1.4 | 2.48 | 269.1 | -2.76 | K0 |
| 258 | 234 | 3 | 44 | 17.3 | 32 | 0 | 15 | 20.59 | 12 | 2710 | 0.013 | 0.67 | | -2.82 | M8 |
| 143 | 139 | 3 | 44 | 17.8 | 32 | 4 | 48 | 18.26 | 5 | 3091 | 0.15 | 2.77 | 4.4 | -2.47 | M5,25 |
| 84 | 83 | 3 | 44 | 17.9 | 32 | 12 | 20 | 15.35 | 20 | 3488 | 0.31 | 1.77 | 165.5 | -2.79 | M2,5 |
| 171 | 163 | 3 | 44 | 18.1 | 32 | 10 | 53 | 19.88 | 16 | 3741 | 0.021 | 4.54 | 282.2 | -1.34 | M0,75 |
| 155 | 149 | 3 | 44 | 18.2 | 32 | 9 | 59 | 18.5 | 8 | 3234 | 0.15 | 3.05 | | -1.88 | M4,25 |
| 29 | 29 | 3 | 44 | 18.2 | 32 | 4 | 57 | 17.34 | 4 | 5945 | 9.6 | 11.84 | 94.6 | -1.65 | G1 |
| 211 | 195 | 3 | 44 | 18.3 | 32 | 7 | 33 | 18.26 | 5 | 3161 | 0.078 | 2.45 | | -2.37 | M4,75 |
| 133 | 130 | 3 | 44 | 18.6 | 32 | 12 | 53 | 18.56 | 115 | 3451 | 0.14 | 4.79 | | -1.03 | M2,75 |
| 91 | 90 | 3 | 44 | 19 | 32 | 7 | 36 | 18.59 | | 3091 | 0.041 | 2.77 | | -2.84 | M5,25 |
| 28 | 28 | 3 | 44 | 19.1 | 32 | 9 | 31 | 11.39 | | 7200 | 6.5 | 1.88 | 6.7 | -2.81 | F0 |
| 90 | 89 | 3 | 44 | 19.2 | 32 | 7 | 35 | 16.72 | 4 | 3306 | 0.26 | 2.62 | 49.4 | -1.53 | M3,75 |
| 251 | 228 | 3 | 44 | 19.2 | 32 | 6 | 0 | 19.16 | 9 | 3024 | 0.038 | 1.56 | | | M5,75 |
| 239 | 217 | 3 | 44 | 19.6 | 32 | 2 | 25 | 19.78 | 9 | 3234 | 0.053 | 4.43 | | -2.24 | M4,25 |
| 304 | 271 | 3 | 44 | 19.6 | 32 | 6 | 46 | 17.77 | 28 | 2880 | 0.009 | 1.74 | | -1.2 | M7 |
| 174 | 166 | 3 | 44 | 20 | 32 | 6 | 46 | 17.72 | 5 | 3342 | 0.12 | 3.19 | 20.4 | | M3,5 |
| 110 | 108 | 3 | 44 | 20.2 | 32 | 8 | 56 | 16.69 | 47 | 3560 | 0.32 | 3.55 | | -1.47 | M2 |
| 230 | 209 | 3 | 44 | 20.3 | 32 | 5 | 44 | 19.31 | 4 | 3161 | 0.075 | 4.15 | | -1.96 | M4,75 |
| 367 | | 3 | 44 | 20.4 | 32 | 1 | 58 | | | | | 0 | | 0.9 | |
| 270 | 243 | 3 | 44 | 21.1 | 32 | 6 | 17 | 20.96 | 8 | 2710 | 0.01 | 0.71 | | -2.58 | M8 |
| 232 | 211 | 3 | 44 | 21.2 | 32 | 1 | 15 | | | 3270 | 0.072 | 10.64 | | -1.8 | M4 |
| 176 | 167 | 3 | 44 | 21.3 | 32 | 12 | 37 | 17.69 | 31 | 3161 | 0.073 | 2.09 | | -1.75 | M4,75 |
| 104 | 102 | 3 | 44 | 21.3 | 32 | 5 | 2 | 16.69 | 30 | 3488 | 0.32 | 3.97 | 194.3 | | M2,5 |
| 111 | 109 | 3 | 44 | 21.3 | 32 | 11 | 56 | 16.53 | | 3560 | 0.27 | 2.8 | 24.3 | -1.83 | M2 |
| | | 3 | 44 | 21.57 | 32 | 15 | 10 | 16.87 | 39 | 3320 | 0.134 | 1.5 | | -1.2 | M5,5 |
| 102 | 101 | 3 | 44 | 21.6 | 32 | 10 | 17 | 16.21 | 2 | 3632 | 0.29 | 2.87 | 313.1 | -2.39 | M1,5 |
| 38 | 38 | 3 | 44 | 21.6 | 32 | 10 | 38 | 16.57 | 48 | 4060 | 0.79 | 5.64 | 32.7 | -1.62 | K7 |
| 109 | 107 | 3 | 44 | 21.7 | 32 | 6 | 25 | 16.23 | 2 | 3451 | 0.29 | 2.77 | 48.4 | -2.6 | M2,75 |
| 153 | 148 | 3 | 44 | 21.8 | 32 | 12 | 31 | 16.96 | 4 | 3342 | 0.12 | 1.84 | 8.5 | -2.53 | M3,5 |
| 86 | 85 | 3 | 44 | 21.9 | 32 | 12 | 12 | 15.96 | 5.5 | 3270 | 0.29 | 2.34 | 14.5 | -2.72 | M4 |
| 361 | | 3 | 44 | 21.9 | 32 | 17 | 27 | | | 3125 | | 0 | | -1.12 | |
| 57 | 57 | 3 | 44 | 22.3 | 32 | 5 | 43 | 16.71 | 34 | 3955 | 0.54 | 4.43 | 72.5 | -1.37 | K8 |
| 92 | 91 | 3 | 44 | 22.3 | 32 | 12 | 1 | 16.43 | 75 | 3705 | 0.33 | 3.19 | 19 | -1.12 | M1 |
| 68 | 67 | 3 | 44 | 22.6 | 32 | 1 | 54 | 16.01 | 3.5 | 3488 | 0.51 | 3.48 | | -2.17 | M2,5 |
| 235 | 214 | 3 | 44 | 22.6 | 32 | 1 | 28 | 21.42 | | 3125 | 0.049 | 5.67 | | -2.63 | M5 |
| 229 | 208 | 3 | 44 | 22.7 | 32 | 1 | 42 | 21.35 | 40 | 3161 | 0.055 | 6.17 | 53.9 | -1.81 | M4,75 |
| 238 | | 3 | 44 | 22.9 | 32 | 14 | 41 | | | 3057 | | 0 | | -1.39 | |
| 105 | 103 | 3 | 44 | 23 | 32 | 11 | 57 | 16.04 | 1.5 | 3524 | 0.22 | 1.88 | 6.4 | | M2,25 |
| 314 | 274 | 3 | 44 | 23 | 32 | 7 | 19 | | 40 | 2990 | 0.002 | 4.04 | | 0.1 | M6 |
| 320 | 276 | 3 | 44 | 23.3 | 32 | 1 | 54 | 19.61 | 13.5 | 3024 | 0.026 | 1.63 | | -2.02 | M5,75 |
| 192 | 179 | 3 | 44 | 23.6 | 32 | 9 | 34 | 17.96 | 5 | 3125 | 0.074 | 1.7 | | -1.51 | M5 |
| 163 | 157 | 3 | 44 | 23.6 | 32 | 1 | 53 | 21.29 | 40 | 3198 | 0.16 | 7.87 | | -1.14 | M4,5 |
| 83 | 82 | 3 | 44 | 23.7 | 32 | 6 | 46 | 15.82 | 3 | 3488 | 0.39 | 2.87 | 46.1 | -2.82 | M2,5 |
| 35 | 35 | 3 | 44 | 24 | 32 | 11 | 0 | 12.67 | | 6030 | 3.1 | 2.59 | 131.9 | -2.72 | G0 |
| 42 | 42 | 3 | 44 | 24.3 | 32 | 10 | 20 | 14.16 | | 4350 | 0.92 | 2.34 | 320.7 | -2.68 | K5 |
| 282 | 253 | 3 | 44 | 24.5 | 32 | 1 | 44 | 21.68 | 200 | 3415 | 0.062 | 8.23 | | -1.33 | M3 |
| | 105 | 3 | 44 | 24.6 | 32 | 3 | 57 | 16.9 | 3 | 3705 | | | | | M1 |
| 224 | 204 | 3 | 44 | 24.6 | 32 | 10 | 3 | 18.14 | | 3198 | 0.04 | 1.6 | | -1.01 | M4,5 |
| 10 | 10 | 3 | 44 | 24.7 | 32 | 10 | 15 | 11 | | 6890 | 13 | 1.91 | 13.2 | -2.86 | F2 |
| 119 | 116 | 3 | 44 | 25.3 | 32 | 10 | 13 | 17.76 | 51 | 3161 | 0.2 | 3.05 | 2.9 | -1.17 | M4,75 |
| 55 | 56 | 3 | 44 | 25.5 | 32 | 11 | 31 | 16.31 | 51 | 3850 | 0.34 | 3.65 | 106.8 | -0.99 | M0 |
| 56 | 55 | 3 | 44 | 25.5 | 32 | 11 | 31 | 16.68 | | | | 0 | | -0.99 | M2,25 |
| 60 | 60 | 3 | 44 | 25.6 | 32 | 12 | 30 | 14.65 | 2.7 | 3778 | 0.45 | 1.28 | 147 | -2.76 | |
| 88 | 87 | 3 | 44 | 25.6 | 32 | 6 | 17 | 18.1 | 3 | 3524 | 0.54 | 6.03 | 30.7 | -1.64 | M0,5 |
| 263 | | 3 | 44 | 25.7 | 32 | 15 | 49 | | | 3057 | | 0 | | -1.31 | |
| 252 | 229 | 3 | 44 | 25.8 | 32 | 9 | 6 | 20.05 | | 3058 | 0.04 | 3.09 | | -1.87 | M5,5 |
| 287 | 258 | 3 | 44 | 25.8 | 32 | 10 | 59 | | | 2990 | 0.006 | 4.89 | | -2.01 | M6 |
| 5 | 5 | 3 | 44 | 26 | 32 | 4 | 30 | 13.79 | 23 | 5520 | 9.9 | 5.32 | 385 | -1.14 | G8 |
| 293 | 264 | 3 | 44 | 26.4 | 32 | 8 | 10 | | | 2400 | 0.002 | 1.6 | | -1.98 | M9 |
| 58 | 58 | 3 | 44 | 26.6 | 32 | 3 | 58 | 16.2 | 11 | 3161 | 0.39 | 1.84 | 150.1 | -2.85 | M4,75 |
| 73 | 72 | 3 | 44 | 26.6 | 32 | 8 | 21 | 17.34 | | | | 0 | | -1.47 | |
| 244 | 221 | 3 | 44 | 26.7 | 32 | 2 | 36 | 19.16 | 27 | 3024 | 0.033 | 1.28 | | -1.6 | M5,75 |
| 72 | 71 | 3 | 44 | 26.7 | 32 | 8 | 20 | 18.01 | 42 | 3778 | 0.53 | 6.42 | 14.4 | -1.47 | M0,5 |
| 290 | 261 | 3 | 44 | 26.9 | 32 | 9 | 26 | | | 2632 | 0.002 | 0.99 | | -2.78 | M8,25 |
| 307 | 273 | 3 | 44 | 26.9 | 32 | 12 | 51 | 21.49 | | 2632 | 0.003 | 0.001 | | -2.32 | M8,25 |
| 65 | 65 | 3 | 44 | 27 | 32 | 4 | 44 | 14.96 | 3 | 3705 | 0.46 | 1.95 | 87.8 | -2.92 | M1 |
| 165 | 159 | 3 | 44 | 27.2 | 32 | 10 | 38 | 18.61 | 26 | 3161 | 0.09 | 3.05 | | -2.01 | M4,75 |
| 298 | 268 | 3 | 44 | 27.2 | 32 | 3 | 47 | | | 2400 | 0.003 | 2.34 | | -2.38 | M9 |
| 113 | | 3 | 44 | 27.2 | 32 | 14 | 21 | 16.53 | 35 | 3342 | | 0 | | -1.42 | |
| 250 | 227 | 3 | 44 | 27.3 | 32 | 7 | 18 | 18.73 | 6 | 3161 | 0.038 | 2.3 | 3.3 | -2.68 | M4,75 |
| 78 | 77 | 3 | 44 | 27.9 | 32 | 7 | 32 | 15.79 | 3 | 3560 | 0.46 | 2.8 | 20.2 | -2.69 | M2 |
| 264 | 238 | 3 | 44 | 28 | 32 | 5 | 20 | 19.28 | 30 | 3058 | 0.021 | 1.24 | | -1.32 | M5,5 |
| 77 | 76 | 3 | 44 | 28.1 | 32 | 16 | 0 | 15.98 | 3 | 3379 | 0.3 | 2.13 | 25 | -2.65 | M3,25 |
| 144 | 140 | 3 | 44 | 28.4 | 32 | 11 | 23 | 21.14 | | 3415 | 0.19 | 8.51 | 141.6 | -2.62 | M3 |
| 62 | 62 | 3 | 44 | 28.5 | 32 | 7 | 22 | 14.67 | 1.4 | 4132 | 0.71 | 2.3 | 63.5 | -2.94 | K6,5 |
| 64 | 64 | 3 | 44 | 28.5 | 31 | 59 | 54 | 14.67 | 5.1 | 3342 | 0.46 | 2.94 | 7.5 | -1.62 | K6,5 |
| 266 | 240 | 3 | 44 | 28.9 | 32 | 4 | 23 | 20.39 | 10 | 3024 | 0.027 | 2.77 | | -1.79 | M5,75 |
| 178 | | 3 | 44 | 28.9 | 32 | 1 | 38 | | | 3370 | | 0 | | -1.04 | |
| 200 | 186 | 3 | 44 | 29.1 | 32 | 7 | 57 | 17.82 | 4 | 3198 | 0.072 | 1.95 | 69.1 | -2.63 | M4,5 |
| 161 | 155 | 3 | 44 | 29.2 | 32 | 1 | 16 | 20.45 | 5 | 3306 | 0.16 | 7.09 | 21.1 | -1.59 | M3,75 |
| 225 | 205 | 3 | 44 | 29.5 | 32 | 4 | 4 | 19.54 | | 3125 | 0.071 | 4.15 | | -2.48 | M5 |
| 37 | 37 | 3 | 44 | 29.7 | 32 | 10 | 40 | 15.54 | 145 | 3955 | 0.76 | 3.58 | 6.6 | -1.44 | K8 |
| 172 | 164 | 3 | 44 | 29.8 | 32 | 0 | 55 | | 140 | 2990 | 0.095 | 2.34 | | -0.99 | M6 |
| 276 | 247 | 3 | 44 | 30 | 32 | 9 | 40 | 21.15 | 50 | 2935 | 0.022 | 2.91 | | -1.07 | M6,5 |
| 241 | 219 | 3 | 44 | 30 | 32 | 8 | 49 | 21.15 | | 2990 | 0.044 | 3.05 | | -1.61 | M6,5 |
| 101 | 100 | 3 | 44 | 30 | 32 | 9 | 21 | 19.6 | | 3488 | 0.45 | 8.12 | 19.3 | -2.56 | M2,5 |
| 175 | | 3 | 44 | 30.1 | 32 | 1 | 18 | | | | | 0 | | -1.81 | |
| 278 | 249 | 3 | 44 | 30.3 | 32 | 11 | 35 | 20.64 | 155 | 3524 | 0.005 | 2.66 | 26.8 | 0.37 | M2,25 |
| 173 | 165 | 3 | 44 | 30.3 | 32 | 7 | 43 | 20.27 | | 3342 | 0.17 | 6.31 | 14.4 | -2.4 | M6 |
| 289 | 260 | 3 | 44 | 30.4 | 32 | 9 | 45 | 18.02 | | 2710 | 0.004 | 1.17 | | -2.35 | M8 |
| 121 | 118 | 3 | 44 | 30.6 | 32 | 6 | 30 | 16.43 | | 3778 | 0.35 | 3.87 | 7.1 | -2.25 | M0,5 |
| 18 | 18 | 3 | 44 | 30.8 | 32 | 9 | 56 | 11.28 | | 8970 | 16 | 2.23 | 6.4 | -0.54 | A2 |
| 217 | 198 | 3 | 44 | 31 | 32 | 5 | 46 | 19.18 | 6 | 3058 | 0.061 | 2.45 | 1.2 | -2.28 | M5,5 |
| 265 | 239 | 3 | 44 | 31 | 32 | 2 | 44 | 21.58 | 60 | 3058 | 0.032 | 5.11 | | -1.43 | M5,5 |
| 189 | | 3 | 44 | 31.1 | 32 | 18 | 49 | | | 3370 | | 0 | | -1 | |
| 186 | 176 | 3 | 44 | 31.2 | 32 | 5 | 59 | 18.28 | 100 | 3778 | 0.097 | 6.91 | | -1.11 | M0,5 |
| 4 | 4 | 3 | 44 | 31.2 | 32 | 6 | 22 | 10.04 | | 7200 | 34 | 1.77 | 14.9 | -2.69 | F0 |
| 233 | 212 | 3 | 44 | 31.3 | 32 | 9 | 29 | 19.21 | 10 | 3415 | 0.14 | 6.38 | 1.8 | -1.95 | M3 |
| 142 | 138 | 3 | 44 | 31.3 | 32 | 10 | 47 | 18.08 | 12 | 3234 | 0.012 | 3.19 | | -1.26 | M4,25 |
| 292 | 263 | 3 | 44 | 31.3 | 32 | 8 | 11 | | 95 | 2990 | 0.002 | 2.09 | | 0.8 | M6 |
| 185 | 175 | 3 | 44 | 31.4 | 32 | 11 | 29 | 18.27 | 3 | 3091 | 0.078 | 1.67 | 7.9 | -1.3 | M5,25 |
| 51 | 51 | 3 | 44 | 31.4 | 32 | 0 | 14 | 20.57 | 38 | 3778 | 0.68 | 9.33 | 168.2 | -1.03 | M0,5 |
| 27 | 27 | 3 | 44 | 31.5 | 32 | 8 | 45 | 13.08 | | 4900 | 2.1 | 2.27 | 91.5 | -2.84 | K2 |
| 201 | 187 | 3 | 44 | 31.7 | 32 | 6 | 53 | 18.57 | 9 | 3058 | 0.086 | 2.2 | 3.3 | | M5,5 |
| 219 | 199 | 3 | 44 | 31.8 | 32 | 12 | 44 | 20.66 | | 3198 | 0.093 | 6.74 | | -2.48 | M4,5 |
| 212 | | 3 | 44 | 31.8 | 32 | 15 | 46 | | | 3125 | 0.092 | 5.43 | | -2.67 | |
| 12 | 12 | 3 | 44 | 32 | 32 | 11 | 44 | 13.79 | | 6030 | 4 | 4.36 | | 1.26 | G0 |
| 13 | 13 | 3 | 44 | 32 | 32 | 11 | 44 | | | 8720 | 29 | 7.52 | | 1.26 | A3 |
| 246 | 223 | 3 | 44 | 32.4 | 32 | 3 | 27 | 19.91 | 140 | 3058 | 0.028 | 2.41 | | -1.4 | M5,5 |
| 67 | 66 | 3 | 44 | 32.6 | 32 | 8 | 56 | 16.23 | 5 | 3415 | 0.47 | 3.37 | 40.5 | -1.95 | M3 |
| 31 | 31 | 3 | 44 | 32.6 | 32 | 8 | 42 | 14.87 | 3 | 3488 | 0.87 | 2.52 | 16.7 | -2.53 | M2,5 |
| 16 | 16 | 3 | 44 | 32.7 | 32 | 8 | 37 | 12.59 | | 5700 | 0.75 | 2.66 | 99.7 | -2.67 | G6 |
| 80 | 79 | 3 | 44 | 32.8 | 32 | 9 | 16 | 16.63 | 4 | 3379 | 0.36 | 3.65 | 67.8 | -2.33 | M3,25 |
| 195 | 182 | 3 | 44 | 32.8 | 32 | 4 | 13 | 17.33 | 30 | 3125 | 0.065 | 1.6 | | | M5 |
| 106 | 104 | 3 | 44 | 33.2 | 32 | 15 | 29 | 15.94 | 2.5 | 3524 | 0.26 | 2.23 | 24.9 | -2.71 | M2,25 |
| 226 | | 3 | 44 | 33.2 | 32 | 12 | 58 | | | 3198 | 0.095 | 8.58 | | -1.45 | |
| 81 | 80 | 3 | 44 | 33.3 | 32 | 9 | 40 | 15.9 | 5 | 3560 | 0.41 | 3.83 | 33.1 | -2.33 | M2 |
| 288 | 259 | 3 | 44 | 33.4 | 32 | 10 | 31 | | 35 | 2555 | 0.004 | 0.39 | | -1.04 | M8,5 |
| 301 | 270 | 3 | 44 | 33.7 | 32 | 5 | 47 | | 70 | 2478 | 0.001 | 0.25 | | -1 | M8,75 |
| 300 | 269 | 3 | 44 | 33.7 | 32 | 5 | 21 | | 90 | 2990 | 0.001 | 0.5 | | -0.9 | M6 |
| 61 | 61 | 3 | 44 | 34 | 32 | 8 | 54 | 14.89 | 2 | 3850 | 0.56 | 2.16 | 88.1 | -2.37 | M0 |
| 222 | 202 | 3 | 44 | 34.1 | 32 | 6 | 57 | 19.67 | 15 | 2838 | 0.032 | 1.06 | | -1 | M7,25 |
| 213 | | 3 | 44 | 34.1 | 32 | 16 | 36 | 18.43 | no cont | 3234 | 0.082 | 3.62 | | -1.47 | |
| 1 | 1 | 3 | 44 | 34.2 | 32 | 9 | 46 | 8.19 | | 15400 | 605 | 1.91 | | -2.76 | B5 |
| 69 | 68 | 3 | 44 | 34.3 | 32 | 10 | 50 | 16.34 | 3 | 3560 | 0.62 | 4.04 | 26.4 | -2.85 | M2 |
| 170 | | 3 | 44 | 34.3 | 32 | 12 | 41 | | | 3342 | 0.12 | 10.21 | | -0.92 | |
| 166 | 160 | 3 | 44 | 34.4 | 32 | 6 | 25 | 18.52 | 8 | 3058 | 0.12 | 2.55 | 7.8 | -2.43 | M5,5 |
| 107 | | 3 | 44 | 34.6 | 32 | 3 | 57 | | 5 | 3705 | 0.44 | 5 | | -2.71 | |
| 210 | | 3 | 44 | 34.7 | 32 | 16 | 0 | 18.7 | | 3324 | | 0 | | -0.53 | |
| 129 | 126 | 3 | 44 | 34.8 | 32 | 11 | 18 | 16.99 | 4 | 3560 | 0.26 | 3.3 | 9.5 | -2.77 | M2 |
| 45 | 45 | 3 | 44 | 34.9 | 32 | 6 | 34 | 14.63 | 1.5 | 4278 | 0.99 | 3.01 | 114.8 | -2.65 | K5,5 |
| 87 | 86 | 3 | 44 | 34.9 | 32 | 9 | 54 | 15.74 | 4 | 3342 | 0.31 | 3.19 | 8.1 | -2.75 | M3,5 |
| 308 | | 3 | 44 | 34.9 | 32 | 15 | 0 | | | | | 0.113 | | -2.06 | |
| 23 | 23 | 3 | 44 | 35 | 32 | 7 | 37 | 14.19 | 2 | 4132 | 1.5 | 2.84 | 136.3 | -2.2 | K6,5 |
| 261 | 236 | 3 | 44 | 35 | 32 | 8 | 57 | 18.08 | 5 | 3161 | 0.032 | 2.94 | | -1.63 | M4,75 |
| 152 | | 3 | 44 | 35 | 32 | 15 | 31 | 17.24 | | 3342 | | 0 | 65.5 | -1.3 | |
| 24 | 24 | 3 | 44 | 35.4 | 32 | 7 | 36 | 16.03 | 32 | 3850 | 0.69 | 3.79 | 79 | -1.12 | M0 |
| 348 | | 3 | 44 | 35.4 | 32 | 7 | 36 | | | | | 0 | | -1.12 | |
| 2 | 2 | 3 | 44 | 35.4 | 32 | 10 | 4 | 9.69 | | 8970 | 137 | 3.83 | 1053 | -1.41 | A2 |
| 139 | 135 | 3 | 44 | 35.5 | 32 | 8 | 56 | 18.67 | 50 | 3091 | 0.16 | 3.62 | | -0.98 | M5,25 |
| 188 | | 3 | 44 | 35.5 | 32 | 8 | 5 | 18.69 | | 3091 | 0.094 | 2.87 | 1.9 | | |
| 203 | 189 | 3 | 44 | 35.7 | 32 | 4 | 53 | 18.45 | 11 | 3024 | 0.056 | 0.78 | 6.9 | -1.78 | M5,75 |
| 120 | 178 | 3 | 44 | 35.7 | 32 | 3 | 4 | 18.1 | 5 | 3379 | 0.13 | 3.33 | | -0.92 | M5,25 |
| | 117 | 3 | 44 | 35.7 | 32 | 3 | 3.6 | 17.53 | | 3091 | | | | | M3,25 |
| 284 | 255 | 3 | 44 | 35.9 | 32 | 11 | 18 | 21.02 | 100 | 2962 | 0.01 | 1.99 | | -1.71 | M6,25 |
| 154 | | 3 | 44 | 35.9 | 32 | 15 | 53 | 16.59 | | 3488 | 0.15 | 2.13 | 11.9 | | |
| 199 | 185 | 3 | 44 | 36 | 32 | 9 | 24 | | 30 | 3091 | 0.07 | 2.77 | 1.4 | -1.09 | M5,25 |
| 315 | | 3 | 44 | 36.2 | 32 | 13 | 4 | | | 2990 | 0.015 | 8.16 | | -1.87 | |
| 295 | 265 | 3 | 44 | 36.4 | 32 | 3 | 5 | | 40 | 2478 | 0.003 | 0.5 | | -0.82 | M8,75 |
| 297 | 267 | 3 | 44 | 36.6 | 32 | 3 | 44 | | 140 | 2710 | 0.004 | 1.74 | | -0.51 | M8 |
| 6 | 6 | 3 | 44 | 36.9 | 32 | 6 | 45 | 11.83 | | 5830 | 17 | 3.9 | 1961 | -2.02 | G3 |
| 127 | 124 | 3 | 44 | 37 | 32 | 8 | 34 | 18.02 | 8 | 3161 | 0.18 | 3.44 | 5.9 | -1.65 | M4,75 |
| 100 | 99 | 3 | 44 | 37.2 | 32 | 9 | 16 | 15.63 | | 4205 | 0.32 | 2.73 | 16 | -1.95 | K6 |
| 198 | 184 | 3 | 44 | 37.3 | 32 | 7 | 11 | 19.46 | | 3161 | 0.12 | 4.93 | 1.4 | -1.63 | M4,75 |
| 97 | 96 | 3 | 44 | 37.4 | 32 | 12 | 24 | 18.1 | 10 | 3560 | 0.34 | 5.25 | 24.1 | -1.5 | M2 |
| 75 | 74 | 3 | 44 | 37.4 | 32 | 6 | 12 | 15.18 | | 4060 | 0.51 | 2.52 | 29.2 | -2.39 | K7 |
| 76 | 75 | 3 | 44 | 37.4 | 32 | 9 | 1 | 16.28 | 7 | 3705 | 0.51 | 4.29 | 56.8 | -0.77 | M1 |
| 291 | 262 | 3 | 44 | 37.6 | 32 | 8 | 33 | | 300 | 3058 | 0.001 | 3.3 | | 0.88 | M5,5 |
| 273 | | 3 | 44 | 37.6 | 32 | 11 | 56 | | | 3270 | 0.057 | 12.27 | | -2.29 | |
| 131 | 128 | 3 | 44 | 37.8 | 32 | 12 | 18 | 18.1 | 6 | 3198 | 0.19 | 3.19 | 26.5 | -2.73 | M4,5 |
| 162 | 156 | 3 | 44 | 37.8 | 32 | 10 | 7 | 19.71 | | 4060 | 0.33 | 8.19 | 14.2 | | K7 |
| 30 | 30 | 3 | 44 | 37.9 | 32 | 8 | 4 | 15.72 | 68 | 4060 | 1.4 | 5 | 1407 | -1.47 | K7 |
| 164 | 158 | 3 | 44 | 38 | 32 | 11 | 37 | 17.77 | 10 | 3270 | 0.12 | 3.05 | | -1.33 | M4 |
| 34 | 34 | 3 | 44 | 38 | 32 | 3 | 30 | 14.57 | 3 | 4205 | 0.99 | 2.8 | 123.1 | -0.98 | K6 |
| 253 | 230 | 3 | 44 | 38.1 | 32 | 10 | 22 | 19.33 | 12 | 2990 | 0.035 | 1.56 | | -2.16 | M6 |
| 123 | 120 | 3 | 44 | 38.4 | 32 | 13 | 0 | 16.32 | 2 | 3850 | 0.21 | 2.2 | 17.7 | | M0 |
| 33 | 33 | 3 | 44 | 38.4 | 32 | 7 | 36 | 14.56 | 2 | 4205 | 1.5 | 3.37 | 256.4 | -0.67 | K6 |
| 53 | 53 | 3 | 44 | 38.5 | 32 | 8 | 1 | 15.73 | 9 | 3669 | 0.72 | 3.69 | 210.1 | -1.66 | M1,25 |
| 89 | 88 | 3 | 44 | 38.6 | 32 | 5 | 6 | 16.95 | 4 | 3270 | 0.33 | 3.3 | 35.9 | -2.53 | M4 |
| 96 | 95 | 3 | 44 | 38.7 | 32 | 8 | 57 | 16.18 | 3 | 3379 | 0.24 | 2.13 | 19 | -2.41 | M3,25 |
| 22 | 22 | 3 | 44 | 38.7 | 32 | 8 | 42 | 16.2 | | 4730 | 4.1 | 6.84 | 922 | -2.84 | K3 |
| 227 | 206 | 3 | 44 | 38.9 | 32 | 6 | 36 | 18.87 | 10 | 2990 | 0.047 | 1.1 | 1.6 | -2.75 | M6 |
| 228 | 207 | 3 | 44 | 39 | 32 | 3 | 20 | 18.3 | 98 | 3125 | 0.046 | 1.84 | | -1.45 | M5 |
| 115 | 112 | 3 | 44 | 39.2 | 32 | 20 | 9 | | 20 | 3198 | 0.2 | 1.95 | | -1.81 | M4,5 |
| 32 | 32 | 3 | 44 | 39.2 | 32 | 7 | 35 | 14.55 | | 4730 | 3.6 | 5.07 | 95.1 | -2.83 | |
| 254 | 231 | 3 | 44 | 39.2 | 32 | 8 | 14 | 20.93 | | 2710 | 0.016 | 1.35 | | -1.88 | M8 |
| 9 | 9 | 3 | 44 | 39.2 | 32 | 9 | 18 | 13.69 | | 5520 | 11 | 5.53 | 573 | -2.76 | G8 |
| 82 | 81 | 3 | 44 | 39.2 | 32 | 9 | 45 | 16.62 | 8 | 3560 | 0.39 | 3.83 | 112.5 | -1.46 | M2 |
| 216 | 197 | 3 | 44 | 39.4 | 32 | 10 | 8 | 18.32 | 5 | 3125 | 0.057 | 1.95 | 1.5 | -2.82 | M5 |
| 71 | 70 | 3 | 44 | 39.8 | 32 | 18 | 4 | 16.85 | 14 | 3306 | 0.28 | 3.12 | 101.6 | -1.8 | M3,75 |
| 54 | 54 | 3 | 44 | 40.1 | 32 | 11 | 34 | 14.8 | | 4900 | 1.4 | 4.22 | 400.2 | -2.69 | K2 |
| 182 | 172 | 3 | 44 | 40.2 | 32 | 9 | 33 | 18.65 | 26 | 3198 | 0.068 | 3.23 | | -0.24 | M4,5 |
| 134 | 131 | 3 | 44 | 40.2 | 32 | 9 | 13 | | | 3125 | 0.22 | 5.14 | 7.8 | -2.17 | M5 |
| 179 | 169 | 3 | 44 | 40.8 | 32 | 13 | 7 | 19 | 9 | 3270 | 0.11 | 3.44 | 13.3 | -2.53 | M4 |
| 221 | 201 | 3 | 44 | 41.1 | 32 | 8 | 7 | 20.57 | | 3091 | 0.071 | 5.11 | | -1.5 | M5,25 |
| 286 | 257 | 3 | 44 | 41.2 | 32 | 6 | 27 | 19.17 | 28 | 3024 | 0.039 | 1.52 | | -1.16 | M5,75 |
| 141 | 137 | 3 | 44 | 41.2 | 32 | 10 | 10 | 18.85 | | 3415 | 0.015 | 5.89 | | -1.89 | M3 |
| 124 | 121 | 3 | 44 | 41.3 | 32 | 10 | 25 | 16.9 | 3 | 3161 | 0.17 | 1.7 | 10.6 | | M4,75 |
| 248 | 225 | 3 | 44 | 41.3 | 32 | 4 | 53 | 19.29 | 7 | 3125 | 0.046 | 3.19 | | -1.98 | M5 |
| 183 | 173 | 3 | 44 | 41.4 | 32 | 13 | 9 | 18.3 | 3 | 3125 | 0.074 | 1.7 | | -2.73 | M5 |
| 280 | 251 | 3 | 44 | 41.6 | 32 | 10 | 39 | 19.95 | 65 | 3024 | 0.014 | 1.31 | | -2.38 | M5,75 |
| 114 | 111 | 3 | 44 | 41.7 | 32 | 12 | 2 | 18.13 | | 3125 | 0.17 | 3.9 | 16.9 | -1.86 | M5 |
| 190 | | 3 | 44 | 41.9 | 32 | 17 | 57 | | | 3198 | 0.088 | 2.52 | | -2.44 | |
| 39 | 39 | 3 | 44 | 42 | 32 | 9 | 0 | 16.83 | 32 | 3234 | 0.37 | 3.3 | | -1.28 | M4,25 |
| 40 | 40 | 3 | 44 | 42.1 | 32 | 9 | 2 | 17.03 | 74 | 3488 | 0.28 | 4.82 | | -1.15 | M2,5 |
| 257 | | 3 | 44 | 42.3 | 32 | 12 | 28 | | | 3125 | 0.08 | 11.31 | | -2 | |
| 125 | 122 | 3 | 44 | 42.6 | 32 | 6 | 19 | 15.12 | 1.5 | 3705 | 0.19 | 0.71 | 30 | -2.47 | M1 |
| 140 | 136 | 3 | 44 | 42.6 | 32 | 10 | 3 | 19.33 | 50 | 3234 | 0.24 | 6.31 | | -1.19 | M4,25 |
| 130 | 127 | 3 | 44 | 42.8 | 32 | 8 | 34 | 18.03 | 91 | 3161 | 0.2 | 4.15 | | -1.27 | M4,75 |
| 180 | 170 | 3 | 44 | 43 | 32 | 10 | 15 | 18.64 | 10 | 3125 | 0.11 | 3.19 | 2.3 | -1.87 | M5 |
| 259 | | 3 | 44 | 43 | 32 | 16 | 0 | | | 3160 | | 0 | | -0.6 | |
| 327 | | 3 | 44 | 43.3 | 32 | 1 | 31 | | | | | 0 | 92.7 | 0.43 | |
| 205 | | 3 | 44 | 43.3 | 32 | 17 | 57 | | | 3270 | 0.11 | 5.21 | | -2.7 | |
| 49 | 49 | 3 | 44 | 43.5 | 32 | 7 | 43 | 16.67 | | 3705 | 0.92 | 5.25 | 70.8 | -2.32 | M1 |
| 256 | 233 | 3 | 44 | 43.7 | 32 | 10 | 48 | 18.5 | 8 | 3161 | 0.024 | 0.85 | 3.1 | -2.63 | M4,75 |
| 70 | 69 | 3 | 44 | 43.8 | 32 | 10 | 30 | 15.95 | | 3669 | 0.62 | 2.98 | 20.1 | -0.48 | M1,25 |
| 331 | | 3 | 44 | 43.9 | 32 | 1 | 38 | | | | | 0 | | -1.66 | |
| 245 | 222 | 3 | 44 | 44.2 | 32 | 8 | 47 | 19.76 | 6 | 3024 | 0.04 | 2.59 | | -1.99 | M5,75 |
| 275 | 246 | 3 | 44 | 44.3 | 32 | 10 | 37 | 20.01 | 7 | 3091 | 0.017 | 2.41 | | -1.69 | M5,25 |
| 181 | 171 | 3 | 44 | 44.6 | 32 | 7 | 30 | 19.3 | 8 | 3091 | 0.11 | 4.08 | 4.6 | -2.39 | M5,25 |
| 94 | 93 | 3 | 44 | 44.6 | 32 | 8 | 12 | 17.2 | 44 | 3560 | 0.38 | 4.75 | 102.8 | -1.08 | M2 |
| 15 | 15 | 3 | 44 | 44.7 | 32 | 4 | 3 | 14.69 | 46 | 3778 | 1.9 | 3.97 | 566.2 | -1.33 | M0,5 |
| 145 | 141 | 3 | 44 | 44.8 | 32 | 11 | 6 | 17.25 | 2.5 | 3451 | 0.17 | 3.16 | 18.4 | -2.7 | M2,75 |
| 99 | 98 | 3 | 44 | 45 | 32 | 13 | 36 | 17.4 | 7 | 3161 | 0.24 | 2.7 | 517.9 | -2.74 | M4,75 |
| 118 | 115 | 3 | 44 | 45.1 | 32 | 14 | 13 | 19.4 | 2 | 3270 | 0.33 | 6.99 | 43.3 | -2.59 | M4 |
| 240 | 218 | 3 | 44 | 45.2 | 32 | 10 | 56 | 19.47 | 25 | 3024 | 0.034 | 1.88 | 4.5 | -0.94 | M5,75 |
| 191 | | 3 | 44 | 45.2 | 32 | 1 | 20 | | | 3023 | | 0 | | -0.46 | |
| 274 | 245 | 3 | 44 | 45.6 | 32 | 11 | 11 | 19.68 | 3 | 3161 | 0.028 | 3.44 | | | M4,75 |
| 269 | | 3 | 44 | 45.6 | 32 | 18 | 20 | | | 3058 | 0.03 | 3.87 | | -1.71 | |
| 277 | 248 | 3 | 44 | 45.9 | 32 | 3 | 57 | 20.66 | | 3024 | 0.025 | 2.87 | | -1.51 | M5,75 |
| 157 | 151 | 3 | 44 | 46.3 | 32 | 11 | 16 | 19.86 | | 3560 | 0.31 | 8.12 | 129.7 | -1.93 | M2 |
| 267 | 241 | 3 | 44 | 46.6 | 32 | 9 | 2 | 21.26 | 11 | 3024 | 0.026 | 3.83 | | -1.39 | M5,75 |
| 135 | 132 | 3 | 44 | 47.6 | 32 | 10 | 56 | 19.21 | 4 | 3234 | 0.22 | 5.74 | 42.5 | -2.93 | M4,25 |
| 17 | 17 | 3 | 44 | 47.7 | 32 | 19 | 12 | | | 8460 | 20 | 5.35 | 371.7 | -2.72 | A4 |
| 151 | 147 | 3 | 44 | 48.8 | 32 | 13 | 22 | 16.5 | 3 | 3451 | 0.13 | 1.81 | 27.9 | -2.56 | M2,75 |
| 214 | | 3 | 44 | 48.8 | 32 | 18 | 46 | | | 3058 | 0.069 | 3.05 | | -1.91 | |
| 148 | 144 | 3 | 44 | 49.8 | 32 | 3 | 34 | 17.17 | 8 | 3198 | 0.13 | 2.09 | 11.6 | -2.74 | M4,5 |
| 302 | | 3 | 44 | 49.9 | 32 | 6 | 14 | | | | | 0 | | 0.32 | |
| 3 | 3 | 3 | 44 | 50.6 | 32 | 19 | 7 | | | 9520 | 135 | 4.47 | | -2.77 | A0 |
| 93 | 92 | 3 | 44 | 51 | 32 | 16 | 9 | 16.29 | 6 | 3379 | 0.35 | 3.12 | 40.7 | -2.67 | M3,25 |
| 193 | 180 | 3 | 44 | 52.1 | 32 | 4 | 47 | | | 3270 | 0.094 | 3.05 | 3.6 | -2.61 | M4 |
| 340 | 282 | 3 | 44 | 52.8 | 32 | 0 | 57 | 16.78 | 3 | 3161 | 0.16 | 1.35 | | -2.63 | M4,75 |
| 156 | 150 | 3 | 44 | 53.7 | 32 | 6 | 52 | 18.04 | 5 | 3270 | 0.18 | 3.9 | 6.6 | -2.89 | M4 |
| 313 | | 3 | 44 | 54.7 | 32 | 4 | 40 | | | | | 0 | | 0.1 | |
| 184 | 174 | 3 | 44 | 55.3 | 32 | 9 | 35 | 18.46 | 7.5 | 3161 | 0.11 | 3.44 | 153.1 | -2.64 | M4,75 |
| 47 | 47 | 3 | 44 | 55.6 | 32 | 9 | 20 | 16.02 | 2 | 4590 | 1.4 | 5.39 | | -2.78 | K4 |
| 209 | 194 | 3 | 44 | 55.9 | 32 | 7 | 27 | 18.36 | 5 | 3161 | 0.075 | 2.45 | | -2.66 | M4,75 |
| 20 | 20 | 3 | 44 | 56.1 | 32 | 9 | 15 | 14.95 | 11 | 5250 | 3.9 | 5.39 | 382.9 | -1.89 | K0 |
| 160 | 154 | 3 | 44 | 56.1 | 32 | 5 | 56 | 16.5 | 5 | 3451 | 0.13 | 1.81 | 9.5 | -2.7 | M2,75 |
| 309 | | 3 | 44 | 56.7 | 32 | 17 | 3 | | | 3125 | 0.013 | 8.62 | | -1.71 | |
| 303 | | 3 | 44 | 57.6 | 32 | 6 | 31 | | | | | 0 | | 0.32 | |
| 236 | 215 | 3 | 44 | 57.7 | 32 | 7 | 42 | 19.66 | 4 | 2935 | 0.033 | 1.24 | | -1.95 | M6,5 |
| 187 | 177 | 3 | 44 | 57.9 | 32 | 4 | 2 | 17.88 | 4.5 | 3091 | 0.071 | 1.18 | 2.8 | -1.67 | M5,25 |
| 310 | | 3 | 44 | 58.4 | 32 | 18 | 12 | | | 2990 | 0.006 | 2.98 | | | |
| 285 | 256 | 3 | 44 | 59.1 | 32 | 10 | 11 | 21.58 | | 3024 | 0.006 | 2.16 | | -2.23 | M5,75 |
| 322 | 277 | 3 | 44 | 59.2 | 32 | 17 | 32 | 18.88 | 5 | 3524 | 0.31 | 6.56 | 80.3 | -2.71 | M2,25 |
| 194 | 181 | 3 | 44 | 59.8 | 32 | 13 | 32 | | 80 | 3198 | 0.086 | 3.19 | | -1.67 | M4,5 |
| 296 | 266 | 3 | 45 | 0.5 | 32 | 3 | 20 | 22.34 | | 3058 | 0.043 | 5.99 | | -1.87 | M5,5 |
| 249 | 226 | 3 | 45 | 0.6 | 32 | 8 | 19 | 18.22 | 5.5 | 3198 | 0.037 | 1.6 | 6 | -1.89 | M4,5 |
| 237 | 216 | 3 | 45 | 1 | 32 | 12 | 22 | | 16 | 3058 | 0.038 | 1.1 | | -1.32 | M5,5 |
| 341 | 283 | 3 | 45 | 1.1 | 32 | 3 | 20 | 17.76 | 5 | 3234 | 0.047 | 1.35 | | -2.76 | M4,25 |
| 317 | 275 | 3 | 45 | 1.1 | 32 | 2 | 26 | 18.96 | 9.5 | 3024 | 0.034 | 1.24 | | -2.61 | M5,75 |
| 25 | 25 | 3 | 45 | 1.4 | 32 | 5 | 2 | | | 8460 | 13 | 3.65 | | -2.63 | A4 |
| 169 | 162 | 3 | 45 | 1.5 | 32 | 12 | 29 | | 5 | 3270 | 0.11 | 1.95 | | -2.63 | M4 |
| 74 | 73 | 3 | 45 | 1.5 | 32 | 10 | 51 | | | 5250 | 0.94 | 2.94 | 68.6 | -2.76 | K0 |
| 326 | 278 | 3 | 45 | 1.6 | 32 | 13 | 17 | | | 3270 | 0.059 | 3.55 | | -2.57 | M4 |
| 36 | 36 | 3 | 45 | 1.8 | 32 | 14 | 28 | | | 4590 | 1.9 | 4.75 | 132.8 | -2.71 | K4 |
| 128 | 125 | 3 | 45 | 2.9 | 32 | 7 | 0 | 17.49 | 8 | 3161 | 0.22 | 2.94 | 7.2 | -2.61 | M4,75 |
| 306 | 272 | 3 | 45 | 3.6 | 32 | 12 | 14 | | | 2632 | 0.002 | 0.001 | | -2.45 | M8,25 |
| 366 | | 3 | 45 | 3.8 | 32 | 0 | 23 | | | | | 0 | | -1 | |
| 272 | 244 | 3 | 45 | 4.1 | 32 | 5 | 5 | 22.27 | 100 | 2880 | 0.006 | 2.7 | | -1.29 | M7 |
| 338 | 280 | 3 | 45 | 4.3 | 32 | 3 | 6 | 18.16 | 3.5 | 3234 | 0.056 | 2.2 | | -2.59 | M4,25 |
| 149 | 145 | 3 | 45 | 4.6 | 32 | 15 | 1 | | | 3234 | 0.17 | 4.04 | | -2.16 | M4,25 |
| 150 | 146 | 3 | 45 | 5.2 | 32 | 9 | 54 | 17.23 | 2 | 3415 | 0.15 | 2.7 | | -2.66 | M3 |
| 281 | 252 | 3 | 45 | 5.3 | 32 | 12 | 16 | 20.42 | 38 | 3058 | 0.022 | 2.87 | | -1.88 | M5,5 |
| 336 | 279 | 3 | 45 | 5.4 | 32 | 3 | 8 | | 25 | 3058 | 0.12 | 2.8 | | -1.75 | M5,5 |
| 339 | 281 | 3 | 45 | 5.8 | 32 | 3 | 8 | 14.59 | 3 | 3850 | 0.54 | 1.7 | 603.5 | -2.65 | M0 |

| | | | | | | | | | | | | | | |
|---|---|---|---|---|---|---|---|---|---|---|---|---|---|---|
| 220 | 200 | 3 | 45 | 6.7 | 32 | 9 | 31 | 18.76 | 6 | 3024 | 0.061 | 1.7 | | -2.65 | M5,75 |
| 19 | 19 | 3 | 45 | 7.6 | 32 | 10 | 28 | | | 5945 | 4.9 | 2.2 | 83.9 | -2.75 | G1 |
| 11 | 11 | 3 | 45 | 8 | 32 | 4 | 2 | | 1.5 | 5800 | 11 | 6.24 | 477.4 | -2.68 | G4 |
| 323 | | 3 | 45 | 13.1 | 32 | 20 | 5 | | | 3125 | | 0 | | -1.42 | |
| 305 | | 3 | 45 | 13.8 | 32 | 12 | 10 | | | 3342 | | 0 | | -0.03 | |
| 337 | | 3 | 45 | 16.3 | 32 | 6 | 20 | | | | | 0 | | -0.36 | |
| 362 | | 3 | 45 | 17.6 | 32 | 7 | 55 | | | | | 0 | | -1.33 | |
| 351 | | 3 | 45 | 17.8 | 32 | 12 | 6 | | | 3306 | | 0 | | -1.86 | |
| 363 | | 3 | 45 | 19.1 | 32 | 13 | 55 | | | | | 0 | | 0.41 | |
| 357 | 287 | 3 | 45 | 20.5 | 32 | 6 | 35 | 14.9 | 11.5 | 3705 | 1.4 | 3.19 | | -1.39 | M1 |
| 355 | | 3 | 45 | 22.2 | 32 | 5 | 45 | | | 2710 | | 0 | | -1.11 | |
| 356 | 286 | 3 | 45 | 25.2 | 32 | 9 | 30 | | 45 | 3306 | 0.4 | 2.62 | | -0.87 | M3,75 |
| 358 | 288 | 3 | 45 | 30.6 | 32 | 1 | 56 | 14.42 | 0.6 | 4205 | 0.71 | 2.34 | | -2.7 | K6 |
| 354 | 285 | 3 | 45 | 32.3 | 32 | 3 | 15 | 16.03 | 7 | 3415 | 0.13 | 1.21 | | -2.78 | M3 |